\documentstyle[12pt]{article}
\begin{document}
\begin{flushright}
UICHEP-TH/93-2
\end{flushright}
{\large {\bf An Alternate Approach to Transition Potentials} }
\begin{center}
Asim Gangopadhyaya$^{(1)}$, Prasanta K. Panigrahi$^{(2)}$, \\
and Uday P. Sukhatme$^{(3)}$
\end{center}
\begin{tabular}{l l}
1) &Department of Physics, Loyola University Chicago, Chicago, IL 60626;\\
2) &Laboratoire de Physique Nucl\'eaire, Universit\'e de Montr\'eal,
    Montr\'eal, \\
   &Qu\'ebec, Canada H3C3J7;\\
3) &Department of Physics, University of Illinois at Chicago, Chicago,
    IL 60680.\\
\end{tabular}
\vspace{1.0in}
\begin{abstract}
We analyze transition potentials
$\left( V(r) \stackrel{r\sim 0}{\rightarrow} {\alpha r^{-2}}\right)$
in non-relativistic quantum mechanics
using the techniques of supersymmetry.
For the range $-{1 \over 4}<\alpha<{3 \over 4}$, the
eigenvalue problem becomes ill-defined (since it is not possible to choose
a unique eigenfunction based on square integrability and boundary conditions).
 It is shown that
supersymmetric quantum mechanics (SUSYQM)
provides a natural prescription for a unique determination of the spectrum.
Interestingly, our SUSYQM based approach picks out the same
"less singular" wave
functions as the conventional approach, and
thus provides a simple justification for the usual practice in the
 literature.  Two examples ( the P\"oschl-Teller II potential
and a two anyon system
on the plane) have been worked out for illustrative purposes.
\end{abstract}
\newpage
\noindent Transition potentials
in non-relativistic quantum mechanics are defined by
$$\lim_{r \rightarrow 0} \; \; r^2 V(r) \; = \; \alpha,$$ with a
finite nonzero $\alpha$\cite{frank}.
For $\alpha>{ 3 \over 4}$, the eigenvalue problem is well defined and
can be solved by conventional means.
For $\alpha<-{ 1 \over 4}$, both the independent solutions of the
Schr\"odinger
differential equation are square integrable at the origin and there is no
mechanism available to select any specific linear combination.

However, for the intermediate range
$-{1 \over 4}<\alpha<{ 3 \over 4}$, transition
potentials exhibit a very interesting behavior.
This range corresponds to the so called
"limit  circle" case in the literature\cite{frank}, and one has to specify
another real number $c={ \lim_{r \rightarrow 0} \;
{{\psi'(r)}\over {\psi(r)}} }$ in order to make the Hamiltonian formally
self-adjoint.
Here, the requirement of square integrability is not sufficient
to determine the eigenvalues.
Even with the stronger
condition that the wave function must vanish at the origin $(c=\infty)$,
the non-uniqueness
still persists, albeit for a smaller range of $\alpha$ given by
$-{1 \over 4}<\alpha<0$.
For values of $\alpha$ in this interval, eigenvalues
are not well defined in the absence of further assumptions.
This lack of uniqueness arises from the fact that both linearly
independent solutions of the Schr\"odinger equation
are well defined near the origin, and the
condition of square integrability does not help us in discarding
one of them.
In such cases, it is customary\cite{frank} to force
the coefficient of the term with smaller power of $r$ to vanish.  This
conventional approach of retaining the "less singular" wave function  then
leads to the determination of eigenvalues, and well defined
eigenfunctions.  Frank et al.\cite{frank}, in their comprehensive study,
justified the above choice through a regularization procedure.
Specifically, the potential is first made regular in a small neighborhood
of the singular point, with a radius $\gamma$.  After matching the solutions
at the boundary and taking the limit $\gamma \rightarrow 0$, the less singular
wave function gets selected.

For any spherically symmetric potential in three dimensions,
the ${1 \over {r^2}}$-term arises from the angular momentum term in the
Hamiltonian.  The requirement of single valuedness constrains the
coefficient of $\alpha$ to be positive.
However, in two-dimensional quantum mechanical systems, angular momentum
can take any real value, and thus the problem of indeterminacy
is relevant to the quantum mechanics of anyons\cite{anyon}.
The problem is also relevant for many known physically interesting
one dimensional potentials (Rosen-Morse, Eckart, P\"oschl-Teller, etc.),
which have a $r^{-2}$ behavior at the origin.

In this paper, we provide an alternate way of determining
eigenvalues in the critical range $-{ 1\over 4} < \alpha < {3 \over 4}$.
Our method is based on the supersymmetric approach to
quantum mechanics\cite{SUSY_QM} (SUSYQM).  The key idea is that
in situations where the eigenfunctions of a potential $V_-$ are not
unambiguously
determined, the supersymmetric partner potential $V_+$ has no
such problem.  Thus, solving for $V_+$ first and then using the
degeneracy relation, one can solve the eigenvalue problem for the
potential $V_-$.  Interestingly, we find that our approach
leads to the same answer as the one stated in Ref.\cite{frank}.
Hence, this paper provides an alternate  justification for the
prescription of choosing the "less singular" solution, which obviously works.
Alternatively, it can be viewed that supersymmetry is properly
realized in a space of eigenstates containing "less singular"
wave functions.

To be complete, we have given a brief review of
SUSYQM.  For a detailed description of SUSYQM, we refer the reader to
Ref.\cite{dutt} and references therein.
We use examples of the P\"{o}schl-Teller potential,
and that of a two anyon system to describe how
SUSYQM provides a method to resolve the indeterminacy
mentioned earlier.
\bigskip

SUSYQM is characterized by a superpotential $W$ and a pair of linear
operators $A$ and $A^{\dagger}$:
\begin{equation}
A=\frac{d}{dr}+W(r),
{}~~A^{\dagger}=-\frac{d}{dr}+W(r).
\label{Aops}
\end{equation}
Combining these operators, we can define two Hamiltonians,
\begin{equation}
H_-=A^{\dagger}
A=-\frac{d^2}{dr^2}+V_-(r),\nonumber
\end{equation}
\begin{equation}
H_+=AA^{\dagger}=-\frac{d^2}{dr^2}+V_+(r),\nonumber
\end{equation}
\begin{equation}
V_{\pm}(r)=W^2(r)\pm  W'(r).
\label{vpm}
\end{equation}
We have set ${\hbar}={2m}=1$.
The potentials V$_+$ and V$_-$ are called supersymmetric partner
potentials. The eigenstates of the Hamiltonians $H_{(-)}$ and $H_{(+)}$ are
$\psi_n^{(-)}$ and $\psi_n^{(+)}$ respectively.
$\psi_n^{(\pm)}$ satisfy the eigenvalue equations
\begin{equation}
H_-\psi_n^{(-)}=E_n^{(-)}\psi_n^{(-)}, ~~ H_+\psi_n^{(+)}=
E_n^{(+)}\psi_n^{(+)}.
\label{eigveq}
\end{equation}
If the ground state of $H_-$ has zero energy, i.e. $E_0^{(-)}=0$,
then supersymmetry is said to be unbroken  and one has $A\,\psi_0^{(-)}=0$.
It then follows from eq.~(\ref{Aops})   that
\begin{equation}
\psi_0^{(-)}=exp~[-\int^rW(r')dr'].
\label{psi0}
\end{equation}
For unbroken supersymmetry, one needs $\psi_0^{(-)}$ or
${ 1 \over {\psi_0^{(-)}} }$
to be an acceptable wave function, i.e. it must be quadratically
integrable  and satisfy correct boundary conditions.  For a finite
domain, the wave function must vanish at the end points.
For a normalizable well-defined $\psi_0^{(-)}$,
one gets the energy degeneracy relation
\begin{equation}
E_{n+1}^{(-)} ~~=~~E_n^{(+)}.
\label{sedeg}
\end{equation}
The corresponding eigenfunctions of $H_-$ and $H_+$ are related by
\begin{equation}
\psi_{n+1}^{(-)} ~~=A^\dagger~\psi_n^{(+)}.
\label{eq8}
\end{equation}

The applicability of SUSYQM to lift the ambiguity
in the determination of eigenvalues and eigenfunctions that plague
transition potentials can be appreciated from the following discussion.
If the superpotential $W(r)$
is given by $-{{l+1}\over r } $
near $r=0$, then potentials $V_-(r)$ and $V_+(r)$
 are of
the form ${ {l(l+1)} \over r^2}$  and ${ {(l+1)(l+2)} \over r^2}$ near $r=0$.
Their wave functions are given by linear combinations
$\left[c_1\;r^{-l} \left(1+{\cal O}(r)\right)\, +
\,c_2\;r^{l+1}\left(1+{\cal O}(r)\right)\,\right]$ and
$\left[c_1'\;r^{-l-1} \left(1+{\cal O}(r)\right)\, +
\,c_2'\;r^{l+2}\left(1+{\cal O}(r)\right)\,\right]$
respectively\cite{tom}.
For $-{1 \over 2} < l < {1 \over 2}$, which corresponds to the
problematic range $-{1\over 4}<\alpha<{3\over 4}$ for $V_-$,
both solutions are
square integrable.  For the potential $V_+$
one has the range $-{1 \over 2} < l < {1 \over 2}$, which is well defined and
one can determine a unique wave function for $V_+$.
A proper wave function (and from it the eigenvalues)
for $V_-$ is obtained by applying the
operator $A^\dagger$ on the solution of $V_+$ as is shown in eq.(8). In what
follows,
we will use two examples to explicitly describe the working of
our approach.

\noindent {\bf Examples:  }

\noindent {\bf (a)~~~P\"oschl-Teller Potential}

\noindent Let us consider the P\"oschl-Teller II superpotential
\begin{equation}
W=A\, \tanh\,r \;-\;B\, {\rm coth}\,r\;\;\;\;\;  (0<r<\infty).
\end{equation}
For $A\,>\,B$ the above superpotential corresponds to a case of unbroken
SUSY. The corresponding supersymmetric partner potentials are
given by,
\begin{eqnarray}
V_-(r) &=& -A(A+1)\, {\rm sech}^2\,r \;+\;
              B(B-1)\, {\rm cosech}^2\,r \;+\;
              \left( A-B \right)^2 \nonumber \\
V_+(r) &=& -A(A-1)\, {\rm sech}^2\,r \;+\;
              B(B+1)\, {\rm cosech}^2\,r \;+\;
              \left( A-B \right)^2.
\end{eqnarray}
Without loss of generality we will assume
$-{1 \over 2}<A<\infty$ and
${1 \over 2} <B < \infty$.
One should note here that $r$-dependent parts of the above potentials are such
that one could obtain $V_+(r)$ from $V_-(r)$ simply by replacing
$A$ by $A-1$, and $B$ by $B+1$. To clearly see the ambiguity in the eigenvalue
problem, we proceed with the analysis of the Schr\"odinger eqation.
The time independent Schr\"odinger equation for $V_-(r)$ is given by
\begin{equation}
{ {d^2\psi^{(-)}} \over {dr^2} } +
\left[ E + { {A(A+1)} \over { {\rm cosh}^2\,r} } -
           { {B(B-1)} \over { {\rm sinh}^2\,r} } -
              \left( A-B \right)^2
\right] \psi^{(-)}(r) =0.
\label{TISE2}
\end{equation}
With a change of variables $y\,=\,{\rm cosh}^2\,r$ and
$\psi(y)\,=\,y^{-{1 \over 2} A} \left( y-1 \right)^{{1 \over 2} B} v(y)$,
and also replacing  $y$ by $1-y$,
eq.(\ref{TISE2}) can be cast in the form of a hypergeometric
equation, i.e.
\begin{equation}
y(1-y) v'' + \left[ \left({1 \over 2} -A \right)
-\left(1-A+B \right)\,y \right]\;v'
-{1 \over 4} E\, v \,=\, 0.
\end{equation}
The general solution is:
\begin{eqnarray}
\psi^{(-)}(r)&=& {\rm cosh}^{-A}r\,{\rm sinh}^{B}r
\left[
c_1 F \left( a',b',c';\,-{\rm sinh}^2r \right) \right. \nonumber \\
&+&
\left.  c_2 {\rm sinh}^{(1-2B)}r
F \left( a'+1-c', b'+1-c', 2-c';-{\rm sinh}^2r \right)\;\right],
\end{eqnarray}
where the constants $a',\,b',\, {\rm and}\, c'$ are given by
%
\begin{eqnarray}
a'&=&{1 \over 2}\left(B-A+Q\right),\nonumber\\
b'&=&{1 \over 2}\left(B-A-Q\right), \nonumber \\
c'&=&B+{1 \over 2} \;\;{\rm  \;\;\;\;\;\;\;\;\; and }\\
Q&=&\sqrt{\left(B-A\right)^2-E} \nonumber\;.
\label{constants1}
\end{eqnarray}
Near the point $r\sim 0$, the solution reduces to
\begin{equation}
\psi^{(-)} \stackrel{r\sim 0}{\rightarrow}
\left[c_1\;r^B \left(1+{\cal O}(r)\right)\, +
\,c_2\;r^{(1-B)}\left(1+{\cal O}(r)\right)\,\right].
\label{BC1}
\end{equation}
Normalizability requires that $\psi^{(-)}$ be less singular than
${1 \over {\sqrt{r}}}$ near origin.
Thus for $B \geq {3 \over 2}$ the wave function becomes non-normalizable
unless $c_2 = 0$.  With $c_2 = 0$, a
subsequent constraint coming from the
requirement of the vanishing of the wave function at infinity
(which is demanded by the normalizability)
suffices to determine eigenvalue $E$ in terms of the parameters
$A \;{\rm and }\; B$.  However, if
$\left({1 \over 2}<B<{3\over 2}\right)$,
both terms on the right
hand side of eq.(\ref{BC1}) are normalizable and hence no
condition is imposed upon the coefficients.
In such cases, we solve the eigenvalue problem for $V_+$ instead,
which has a well defined set of eigenfunctions and eigenvalues.  Then
using the operators $A$ and $A^\dagger$, we can determine eigenfunctions
and eigenvalues of $V_-$. As an explicit example, let us choose
$B={3\over 4}$ and determine corresponding eigenvalues of the Hamiltonian
$H_+$.
The wave function for the Hamiltonian $H_-$ has the following form near the
origin:
\begin{equation}
\psi^{(-)} \stackrel{r \sim 0}{\rightarrow}
\left[ c_1 r^{3 \over 4} \left(1+{\cal O}(r)\right)\, +
c_2 r^{1 \over 4} \left(1+{\cal O}(r)\right)\, \right].
\label{psi_}
\end{equation}
We see that both the terms of eq.(\ref{psi_}) are well defined near
$r \sim 0$, and hence no constraints are placed on their coefficients
from requiring normalizability around the origin.
The wave function for the Hamiltonian $H_+$ is given by
\begin{equation}
\psi^{(+)} \stackrel{r\sim 0}{\rightarrow} \left[c_1\;
r^{\left({7\over 4}\right)}\left(1+{\cal O}(r)\right)\,
\,+\,c_2\;r^{\left(-{3\over 4}\right)}\left(1+{\cal O}(r)\right)\,\right].
\label{BC2}
\end{equation}
Clearly, normalizability of $\psi^{(+)}$ requires that we set $c_2=0$.
To determine the eigenvalues of $H_+$, we have to study the behavior
at infinity, and for that one uses an alternate
asymptotic form of the hypergeometric function:
\begin{equation}
F(a,b,c;z) \stackrel{z\sim \infty}{\rightarrow}
{ {\Gamma(c)\Gamma(b-a)} \over {\Gamma(b)\Gamma(c-a)} }
(-z)^{-a} \,+\,
{ {\Gamma(c)\Gamma(a-b)} \over {\Gamma(a)\Gamma(c-b)} }
(-z)^{-b} .
\end{equation}
This leads to
\begin{equation}
\psi^{(+)} \stackrel{r \sim \infty}{\rightarrow}
	{ {\Gamma(c)\Gamma(b-a)} \over {\Gamma(b)\Gamma(c-a)} }
	e^{(-Qr)} +
        { {\Gamma(c)\Gamma(a-b)} \over {\Gamma(a)\Gamma(c-b)} }
	e^{(+Qr)}
\label{asymptote}
\end{equation}
where the constants $a,\;b,\; {\rm and}\; c$ are given by
\begin{eqnarray}
a&=&{1 \over 2}\left(B-A+2+Q\right),\nonumber\\
b&=&{1 \over 2}\left(B-A+2-Q\right),\;\;{\rm and }
\nonumber \\
c&=&B+{1 \over 2} .
\label{constants2}
\end{eqnarray}
The second term on the R.H.S. of
eq.(\ref{asymptote}) must vanish to have a well defined bound
state.  This can be achieved if $a$ or
$(c-b)$ is equal to a negative number (say $-k$).  If
$a=-k$, then the eigenvalues are
given by
$$E^{(-)}_k = \left( A-B \right)^2 - \left[ A-B-2k-2 \right]^2
\;\;\;k=0,\,1,\, \cdots ,n.$$
The integer $n$ gives the number of bound states that the potential will
hold, and is related to the parameters $A$ and $B$.  It is the largest
integer satisfying $A-B-2>2n$.  Under this condition, one can show that
$exp~[-\int^rW(r')dr']$ is a well defined function, and hence we have
a supersymmetric situation.  The eigenvalues for the Hamiltonian $H_-$
will be the same as that for $H_+$, except that $H_-$ will have an
additional state (ground state) with zero energy.
The eigenfunctions of $H_+$ are given by\cite{de}
$$\psi^{(+)}(r) \;=\; \left( {\rm sinh\,r}\right)^{(1+ {3\over 4})}
\;\left({\rm cosh\,r}\right)^{-(A-1)}\;
P_k^{\left(1+{3\over 4}-{1 \over 2},\,-(A-1)-{1 \over 2}\right)}
\left({\rm cosh}2r\right).$$
Now the eigenfunctions of the Hamiltonian $H_-$ will be given by
applying the operator $A^\dagger$ [defined in eq.(\ref{Aops})]
on the function $\psi^{(+)}$.  Near the origin $\psi^{(+)}$ is given by
$$\psi^{(+)} \stackrel{r \sim 0}{\rightarrow} r^{7 \over 4}.$$  Now
operating $A^\dagger$ on $\psi^{(+)}$ lowers the power of $r$ by unity,
and hence
$$\psi^{(-)} \stackrel{r \sim 0}{\rightarrow} r^{3 \over 4}.$$
Comparing this above expression with eq.(\ref{psi_}), we see that
SUSYQM automatically chooses the term with higher power of $r$,
which is consistent with the prescription of Ref.\cite{frank}.
Hence the eigenvalues obtained will also be the same.
Thus this method provides a justification for the usual practice of dropping
the term with lower power of $r$ in case of ambiguity.

Instead, if the second condition holds i.e. $c-b = -k$,
the eigenvalues are given by
$E^{(-)}_k = \left( A-B \right)^2 - \left[ A+B+2k+1 \right]^2
\;\;\;k=0,\,1,\, \cdots ,n$.
The condition on $A$ for $n$-bound states in the second case is
given by $A<-B-2n-1$, which can not be satisfied as we have assumed
(without loss of generality) $-{1 \over 2}<A<\infty$ and
${1 \over 2} < B < \infty$.

\medskip

\noindent{\bf (b)~~~Anyons in a Spherically Symmetric Potential}

\medskip

\noindent  Here we consider a system of two anyons and proceed
along similar lines as above. The motion can be divided into center
of mass motion and the dynamics of the relative coordinate.
The two body Hamiltonian is given by\cite{anyon,roy}
\begin{eqnarray}
H &=& -{1 \over {2m} }
\left\{ \partial_{1i} - e\;A_i(\vec{r}_1 - \vec{r}_2) \right\}^2
\nonumber\\
&& -{1 \over {2m} }
\left\{ \partial_{2i} - e\;A_i(\vec{r}_2 - \vec{r}_1) \right\}^2 \; +  \;
V\left(|\vec{r}_1 - \vec{r}_2|\right).
\end{eqnarray}
The vector potential is
\begin{equation}
A_i \;=\; { \Phi \over {2\pi} } {  {\epsilon_{i\;j} r^j} \over {r^2}  };
\;\;\;\Phi \;=\; {\theta \over e},
\end{equation}
where $\theta$ is the well known statistics parameter.
Now defining the center-of-mass and relative coordinates as
$$\vec{R}\;=\; {1 \over 2}(\vec{r}_1 + \vec{r}_2),\;
\vec{r} \;=\; (\vec{r}_1 - \vec{r}_2),$$
the Hamiltonian can be written as
\begin{eqnarray}
H_{c.m} &=& -{ 1 \over {4m} } \nabla^2_R \\
H_{rel} &=&     -{1 \over {2\mu} }
\left\{ \partial_{i} - e\;A_i \right\}^2
+V\left(|\vec{r}|\right),
\end{eqnarray}
where $\mu$ represents the reduced mass of the system.
The radial part of the Hamiltonian for the
relative coordinate is given by
\begin{equation}
H_{rel} \;=\; -{ 1 \over {2\mu} } \left[
               { {d^2} \over {dr^2} } +
               { 1 \over {r} }{ {d} \over {dr} } +
               { 1 \over {r^2} } \left( m +
               {\theta \over \pi} \right)^2 \right] +
               V\left(r\right).
\end{equation}
The Schr\"{o}dinger equation for the relative coordinate
is then given by
\begin{equation}
              - \left[ { {d^2\;\psi} \over {dr^2} } +
               { 1 \over {r} }{ {d\;\psi} \over {dr} } +
               { 1 \over {r^2} } \left( m +
               {\theta \over \pi} \right)^2 \;\psi \right] +
               \left( V\left(r\right) - E \right)\psi\;=\;0,
\end{equation}
where we have set ${2\mu} \;=\;1$.
Substituting $\psi~=~{\phi \over {\sqrt{r}} }$, we get
\begin{equation}
-\phi ''\;+\; \left[ { {(\nu^2-{1 \over 4})} \over {r^2} }\;-\;
E \right]\phi \;=\;0,
\label{reduced}
\end{equation}
where
$\nu\;=\;{ \left( m + {\theta \over \pi} \right)^2 \over {r^2} }$.
Eq.(\ref{reduced}) can now be interpreted as an one
dimensional equation where the domain of the variable $r$ is
given by $0 \leq r <\infty$.  Now the important question is the
boundary condition.  This question has been recently analyzed by Roy and
Tarrach,\cite{roy} who conclude that more general  boundary conditions
$\phi'(0)= c \phi(0)$, $c\neq 0,\infty$ are not allowed
because the boundary condition break supersymmetry. It is interesting to
observe that in the three-dimensional monopole problem, the above mentioned
boundary conditions are also extremely important\cite{luc}.
If we ask for the overlap of the two particles
to be zero, we require $\psi\rightarrow 0$ as
$r\rightarrow 0$. This is equivalent to saying that the
configuration space has been reduced to
$R^2\times (R^2-\{0\}) /Z_2$.  For $\phi$, that would imply
that it goes to zero faster than $\sqrt{r}$.

However, if we
only stipulate a need of square integrability, it implies
that $\phi$ be less singular than
${1 \over \sqrt{r}}$.  The solutions of eq.(\ref{reduced})
are then of the form $r^{l+1}$ and $r^{-l}$, where $l$ is given by
$l(l+1) = {(\nu^2-{1 \over 4})}$.  One can show that for
$l>{1 \over 2}$ or $l<-{1 \over 2}$, only one of the above two
solutions is square integrable near the origin, and this leads
to the unambiguous determination of eigenvalues and eigenfunctions.
However, if   $-{1 \over 2}<l<{1 \over 2}$ then both solutions vanish at
the origin,
and are also square integrable.  Hence it is not possible to choose one over
the other.  For the anyon problem
one starts with a superpotential of the form \cite{roy}
$W(r)= {l+1\over r} + f(r)$, where f(r) has to be suitably chosen to give
the required spherically symmetric potential. One can easily see that in the
partner Hamiltonian the singular term is of the form ${(l+1)(l+2)\over r^2}$
and hence just like the P\"oschl-Teller case there will be no ambiguity in this
sector. As has been worked out in the previous example, the application of the
degeneracy theorem will then give the less singular wave function in the
$H_-$ sector.  Thus, we find that our SUSYQM based formalism gives a clear
cut way of finding the eigenvalues and eigenfunctions of transition
potentials in the region of ambiguity.  Also, the eigenfunctions turn
out to be the same "less singular" type that is commonly chosen in the
literature\cite{frank}.

We would like to thank R. Dutt for many relevant discussion.
One of us (A.G.) thanks Profs. C. M. Brodbeck, J. J. Dykla, J. V. Mallow,
and G. P. Ramsey for discussions and the Physics Department
of the University of Illinois at Chicago for warm hospitality.
P.K.P acknowledges discussion with V. Spiridonov.
This work was supported in part by the U.S. Department of
Energy under grant number DE-FG02-84ER40173 and by the National
Science and Engineering Research Council (NSERC) of Canada.


\newpage

\begin{thebibliography}{99}
\bibitem{frank}	W. M. Frank, D. J. Land, and R.M. Spector,
		Rev. of Mod. Phys. {\bf 43}, 36 (1971).
\bibitem{anyon}               R. Mackenzie and F. Wilczek, Int. J. Mod. Phys.
                A {\bf 3}, 2877 (1988),\\
C. Manuel and R. Tarrach, Phys. Lett. {\bf B~268}, 222 (1991),\\
               R. Jackiw, CTP Preprint No. 1937 (January, 1991).
\bibitem{SUSY_QM}
%
		E. Witten, Nucl. Phys. {\bf B~185}, 513 (1981);\\
%
		F. Cooper and B. Freedman, Ann. Phys. (N.Y.)
		{\bf 146}, 262 (1983).
%
\bibitem{dutt}	R. Dutt, A. Khare, and U. Sukhatme,
		Am. J. Phys. {\bf 56}, 163 (1988).
\bibitem{tom}   T. D. Imbo and U. P. Sukhatme, Phys. Rev. Lett. {\bf 54},
		2184 (1985).
\bibitem{de}     R. De, R. Dutt, and U. Sukhatme,
		J. Phys. A: Math. Gen. {\bf 25}, L843 (1992).
%
\bibitem{roy}  P. Roy and R. Tarrach, Phys. Lett. {\bf B~274}, 59 (1992),\\
                B. Roy, A. O. Barut, and P. Roy,
		Phys. Lett. {\bf A~172}, 316 (1993).

\bibitem{luc}  L. Vinet and D'Hoker, Comm. Math. Phys. {\bf 97}, 391 (1985).
\end{thebibliography}
\end{document}